\documentclass[12pt,twoside,english]{article}
\usepackage[T1]{fontenc}
\usepackage[latin1]{inputenc}
\usepackage{babel}
\begin{document}

\title{Penetration depth of an electric field in a semi-infinite classical
plasma}

\author{{\normalsize{}M. Apostol }\\
{\normalsize{}Department of Theoretical Physics, Institute of Atomic
Physics, }\\
{\normalsize{}Magurele-Bucharest MG-6, POBox MG-35, Romania }\\
{\normalsize{}email: apoma@theory.nipne.ro}}

\maketitle
\relax
\begin{abstract}
It is shown that the penetration of an oscillating electric field
in a semi-infinite classical plasma obeys the standard exponential
attenuation law $e^{-x/\lambda_{e}}$ (besides oscillations), where
$x$ is the distance from the wall and $\lambda_{e}$ is the extinction
length (penetration depth, attenuation length). The penetration depth
is computed here explicitly; it is shown that it is of the order $\lambda_{e}\simeq[\mid\varepsilon\mid/(1-\varepsilon)]^{1/3}v_{th}/\omega$,
where $\varepsilon$ is the dielectric function, $\omega$ is the
frequency of the field and $v_{th}=\sqrt{T/m}$ is the thermal velocity
($T$ being the temperature and $m$ the particle (electron) mass).
The result is obtained by including explicitly the contribution of
the surface term. 
\end{abstract}
\relax


\section{Introduction}

It is well known that there exists a mechanism of energy transfer
between collective modes and individual particles in collisionless
classical plasmas, governed by the Landau damping.\cite{key-1} The
origin of this mechanism is the causal character of the response of
the plasmas to external excitations. The Landau damping received much
interest, due to its application to heating plasmas by radiofrequency
electric fields.\cite{key-2}-\cite{key-6} Also, the Landau damping
enjoyed controversies along the years, as a consequence of the counter-intuitive
character of an energy loss in collisioness plasmas.\cite{key-7}-\cite{key-19}
Apart from theoretical and experimental investigations, numerical-analysis\cite{key-2}-\cite{key-5}
and mathematical studies are devoted to the phenomenon,\cite{key-20}-\cite{key-23}
which show both the complexity of the concept and difficulties related
to its understanding at the fundamental level.

In semi-infinite plasmas the Landau damping appears as attenuated
spatial oscillations (vibrations). This phenomenon, with its characteristic
penetration depth, has a particular relevance for surface effects.
Specifically, the Landau damping in semi-infinite plasmas implies
an attenuated electric field, with spatial and temporal oscillations,
besides a uniform component, as a response to a uniform oscillating
external electric field perpendicularly applied to the plasma surface.
The calculation of the exact form of this response is complicate,
due, on one hand, to the difficulties related to the Landau damping,
and, on the other hand, as a consequence of the presence of the surface.
The latter point is particularly interesting, because the response
is discontinuous at the surface, and the usual Fourier or Laplace
techniques may not include properly this discontinuity. In addition,
the surface boundary conditions may bring further complications. These
difficulties have been analyzed recently in a clear formulation in
Ref. \cite{key-26}. In various approximations (see, for instance,
Refs. \cite{key-24}-\cite{key-27}), including the original calculation
in Ref. \cite{key-1}, the asymptotically attenuated field is presented
as being proportional to $x^{2/3}e^{-\frac{3}{4}(\omega x/v_{th})^{2/3}}$,
where $x$ is the distance from the wall, $\omega$ is the frequency
of the field and $v_{th}=\sqrt{T/m}$ is the thermal velocity, $T$
being the temperature and $m$ being the particle mass (electrons);
sometimes, an exponential attenutation $\sim e^{-\omega_{0}x/v_{th}}$
is included, where $\omega_{0}$ is the plasma frequency. A non-linear
$x$-dependence ($\sim x^{2/3})$ is related to model assumptions
made upon the surface and an asymptotic treatment of the Landau damping
for the Boltzmann kinetic equation (see, for instance, Ref. \cite{key-26}).
We show here that, when the surface condition (surface term) is included
explicitly, the attenuated field obeys the standard exponential attenuation
law $e^{-x/\lambda_{e}}$ (apart from factors oscillating in space),
where $\lambda_{e}$ is an extinction length (penetration depth, attenuation
length) which is computed here explicitly; up to immaterial numerical
factors, it is of the order $\lambda_{e}\simeq[\mid\varepsilon\mid/(1-\varepsilon)]^{1/3}v_{th}/\omega$,
where $\varepsilon$ is the dielectric function.

\section{Semi-infinite plasma}

We consider a classical plasma at thermal equlibrium consisting of
mobile charges $q$ with mass $m$ and concentration $n$ (electrons)
moving in a rigid neutralizing background. We confine this plasma
to a semi-infinite space (half-space) $x>0$, bounded by a plane surface
$x=0$. The plasma is subject to a uniform oscillating external electric
field $E_{0}e^{-i\omega t}$, where $E_{0}$ is directed along the
$x$-direction (capacitively coupled plasma). The plasma is governed
by the Maxwell distribution. The mean thermal velocity is sufficiently
small to consider plasma unmagnetized. Since the field is directed
along the $x$-direction we may integrate over the transverse velocities
and use for the Maxwell distribution $F=n(\beta m/2\pi)^{1/2}e^{-\frac{1}{2}\beta mv^{2}}$,
where $v$ is the velocity along the $x$-direction and $\beta=1/T$
is the reciprocal temperature. In the collisionless regime the change
$f(x,v)e^{-i\omega t}$ in the Maxwell distribution is governed by
the Boltzmann (Vlasov) equation
\begin{equation}
-i\omega f+v\frac{\partial f}{\partial x}+\frac{q}{m}(E_{0}+E+E_{1})\frac{\partial F}{\partial v}=0\,\,\,,\label{1}
\end{equation}
 where $E$ is a uniform internal electric field and $E_{1}$ is another
internal electric field, which may vary in space; these fields are
generated by internal charges and currents. The uniform reaction field
$E$ occurs in an infinite space too, \emph{i.e.} a space bounded
by surfaces at infinity (it is a bulk reaction field), while the non-uniform
field $E_{1}$ is due to the presence of the surface (it is a surface
field). We seek the solution of equation (\ref{1}) as $f(x,v)=f_{0}(v)+f_{1}(x,v)$,
where 
\begin{equation}
-i\omega f_{0}+\frac{q}{m}(E_{0}+E)\frac{\partial F}{\partial v}=0\label{2}
\end{equation}
 and 
\begin{equation}
-i\omega f_{1}+v\frac{\partial f_{1}}{\partial x}+\frac{q}{m}E_{1}\frac{\partial F}{\partial v}=0\:\:.\label{3}
\end{equation}

The uniform part $f_{0}$ of the solution does not generate charge
density in plasma; it generates a current density. Therefore, it should
satisfy the equation 
\begin{equation}
i\omega E=4\pi q\int dv\cdot vf_{0}\,\,;\label{4}
\end{equation}
it is easy to see that this equation arises from the general equation
$\partial\mathbf{E}/\partial t+4\pi\mathbf{j}=0$, where $\mathbf{j}$
is the current density; this equation ensures the vanishing of the
(internal) magnetic field, as expected. The non-uniform part $f_{1}$
of the solution generates a charge density in plasma; it satisfies
the equation 
\begin{equation}
\frac{\partial E_{1}}{\partial x}=4\pi q\int dvf_{1}\,\,.\label{5}
\end{equation}

The solution of equations (\ref{2}) and (\ref{4}) is 
\begin{equation}
f_{0}=-\frac{iq\omega E_{0}}{m(\omega^{2}-\omega_{0}^{2})}\frac{\partial F}{\partial v}\label{6}
\end{equation}
 and 
\begin{equation}
E=\frac{\omega_{0}^{2}}{\omega^{2}-\omega_{0}^{2}}E_{0}\,\,,\,\,E_{t}=E_{0}+E=\frac{\omega^{2}}{\omega^{2}-\omega_{0}^{2}}E_{0}\,\,\:,\label{7}
\end{equation}
 where $\omega_{0}=(4\pi nq^{2}/m)^{1/2}$ is the plasma frequency;
we recognize here the response of a boundless plasma to an electric
field (restricted to $x>0$), where $\varepsilon=1-\omega_{0}^{2}/\omega^{2}$
is the dielectric function and $E_{t}$ is the total field in plasma
($P=\chi E_{t}$ is the polarization and $\chi=(\varepsilon-1)/4\pi=-nq^{2}/m\omega^{2}$
is the electric susceptibility).

In order to deal conveniently with the boundary condition at the surface
we multiply equation (\ref{3}) by the step function $\theta(x)$
($\theta(x)=1$ for $x>0$, $\theta(x)=0$ for $x<0$) and restrict
ourselves to the solution for $x>0$; equation (\ref{3}) becomes
\begin{equation}
-i\omega f_{1}+v\frac{\partial f_{1}}{\partial x}+\frac{q}{m}E_{1}\frac{\partial F}{\partial v}=vf_{s}\delta(x)\,\,\,,\label{8}
\end{equation}
 where $f_{s}=f_{s}(v)=f_{1}(x=0,v)$; we can check directly this
surface term by integrating equation (\ref{8}) along a small distance
perpendicular to the surface $x=0$. Similarly, equation (\ref{5})
becomes 
\begin{equation}
\frac{\partial E_{1}}{\partial x}-E_{1s}\delta(x)=4\pi q\int dvf_{1}\,\,\,,.\label{9}
\end{equation}
where $E_{1s}=E_{1}(x=0)$. The inclusion of the surface $\delta$-terms
in equations (\ref{8}) and (\ref{9}) is the main point of this paper. 

In equations (\ref{8}) and (\ref{9}) we use the Fourier transforms
with respect to the coordinate $x$ (and restrict ourselves to $x>0$);
we get 
\begin{equation}
f_{1}(k,v)=\frac{i}{\omega-vk+i\gamma}\left[vf_{s}(v)-\frac{q}{m}\frac{\partial F}{\partial v}E_{1}(k)\right]\label{10}
\end{equation}
 and 
\begin{equation}
E_{1}(k)=\frac{4\pi q\int dv\frac{vf_{s}(v)}{\omega-vk+i0^{+}}-iE_{s1}}{k+\frac{4\pi q^{2}}{m}\int dv\frac{\partial F/\partial v}{\omega-vk+i0^{+}}}\,\,\:,\label{11}
\end{equation}
where $\gamma\rightarrow0^{+}$. It is worth noting that in the Fourier
transforms we replace $\omega$ by $\omega+i\gamma$, $\gamma\rightarrow0^{+}$,
in order to ensure the causal behaviour (\emph{i.e.}, zero response
for time $t<0$, which requires a pole in the lower $\omega$-half-plane).
This procedure gives a pole in the upper $k$-half-plane (this is
the connection between the Landau damping and the spatial decay).
At the same time, in the integrals with respect to $v$ we may take
the limit $\gamma\rightarrow0^{+}$, which avoids the singularity
$\omega=vk$; the insertion of the parameter $\gamma$ produces the
Landau damping. We denote by $A$ the denominator in equation (\ref{11});
it can be estimated as
\begin{equation}
\begin{array}{c}
A=k+\frac{4\pi q^{2}}{m}\int dv\frac{\partial F/\partial v}{\omega-vk+i0^{+}}=k+\frac{4\pi q^{2}}{m}P\int dv\frac{\partial F/\partial v}{\omega-vk}-i\frac{4\pi^{2}q^{2}}{mk}\frac{\partial F}{\partial v}\mid_{v=\omega/k}\simeq\\
\\
\simeq k(1-\omega_{0}^{2}/\omega^{2})-i\frac{4\pi^{2}q^{2}}{mk}\frac{\partial F}{\partial v}\mid_{v=\omega/k}\,\,;
\end{array}\label{12}
\end{equation}
we can see that the zeroes of $A$ give the damped collective eigenmodes
$\omega=\pm\omega_{0}-i\Gamma$ (plasma frequency), where $\Gamma$
is given by the imaginary part in equation (\ref{12})\\
 ($\Gamma\simeq-2\pi^{2}q^{2}\omega_{0}/mk^{2})(\partial F/\partial v)\mid_{v=\omega_{0}/k}$);
this is the Landau damping.

\section{Penetrating electric field}

In order to estimate the field $E_{1}(x)$ we need the zeroes of $A$
with respect to $k$ in equation (\ref{11}). It is convenient to
introduce the variable $\xi=\sqrt{\beta m/2}\omega/k$. We can see
easily that the zeroes of $A$ are given by $\xi^{2}\mid\xi\mid e^{-\xi^{2}}=-i\alpha$,
where $\alpha=\mid\varepsilon\mid/2\sqrt{\pi}(1-\varepsilon)$; we
consider the case $\omega<\omega_{0}$ ($\varepsilon<0$; the rather
unrealistic case $\omega>\omega_{0}$ can be treated similarly, by
using the equation $\xi^{2}\mid\xi\mid e^{-\xi^{2}}=i\alpha$). For
small values of $\alpha$ we get two roots of the equation $A=0$,
given by $k_{1,2}\simeq\pm\frac{1}{2\alpha^{1/3}}\sqrt{\beta m}\omega(1+i)$;
only $k_{1}$ (placed in the upper half-plane) contributes to the
$k$-integration for $x>0$. In estimating the integral in the numerator
of equation (\ref{11}) we may leave aside the contribution of the
principal value. For $k$ near $k_{1}$ the field $E_{1}(k)$ has
the form
\begin{equation}
\begin{array}{c}
E_{1}(k)\simeq\frac{B}{k-k_{1}+\frac{i}{5}(k-k_{1})^{*}}\,\,,\\
\\
B=\frac{8\sqrt{2}\pi q\alpha^{2/3}v_{th}^{2}}{5\omega\mid\varepsilon\mid}(1+i)f_{s}\left(\alpha^{1/3}v_{th}(1-i)\right)+\frac{2i}{5\mid\varepsilon\mid}E_{1s}\,\,.
\end{array}\label{13}
\end{equation}
 The reverse Fourier transformation leads to
\begin{equation}
\begin{array}{c}
E_{1}(x)=E_{1s}e^{(i-1)\omega x/2\alpha^{1/3}v_{th}}\end{array}\label{14}
\end{equation}
 with the relationship 
\begin{equation}
E_{1s}=-\frac{8\sqrt{2}\pi q\alpha^{2/3}v_{th}^{2}}{(2+5\mid\varepsilon\mid)\omega}(1-i)f_{s}\left(\alpha^{1/3}v_{th}(1-i)\right)\label{15}
\end{equation}
 (or $E_{1s}=iB$). The final result is given by $E_{1}(t,x)=Re\left[E_{1}(x)e^{-i\omega t}\right]$.
We can see that an additional, non-uniform, electric field $E_{1}(x)$
appears as a result of the presence of the surface. This field oscillates
in space and is attenuated with an attenuation length (penetration
depth, extinction length) $\lambda_{e}\simeq(1/\pi)^{1/6}[\mid\varepsilon\mid/(1-\varepsilon)]^{1/3}v_{th}/\omega$.
It is worth noting that the penetration depth and the wavelength of
the spatial oscillations have the same order of magnitude.

\section{Discussion and conclusions}

Making use of $E_{1}(k)$ given by equations (\ref{11}) and (\ref{13})
we can calculate the change $f_{1}(x,v)$ in the distribution function
(equation (\ref{10})); if we limit ourselves to slow spatial oscillations,
we get 
\begin{equation}
f_{1}(x,v)\simeq-\frac{iq}{m\omega}sgn(v)\frac{\partial F}{\partial v}E_{1}(x)\label{16}
\end{equation}
 (compare with equations (\ref{6}) and (\ref{7})). Within this approximation
$f_{s}(v)=-(iq/m\omega)sgn(v)(\partial F/\partial v)E_{1s}$ and the
polarization charge and current densities are zero (as expected for
slow oscillations).

The amplitude of the field $E_{1}(x)$ depends on the parameter $E_{1s}$,
which accounts for the boundary condition at $x=0$. It is related
to $f_{s}(v)=\frac{1}{2\pi}\int dkf_{1}(k,v)$ by equation (\ref{15}),
where $f_{1}(k,v)$ is given by equation (\ref{10}); it is easy to
see that the integration of the first term in equation (\ref{10})
gives $f_{s}$, while, making use of equations (\ref{13}), the integration
with respect to $k$ of the term which includes $E_{1}(k)$ is zero. 

Within the kinetic approach we may estimate the local change in temperature
by $\delta T=2T\overline{f/F}$, where the overbar implies an integration
over velocities (thermal average). We can see that only $f_{1}$ contributes
to this integration. Making use of equation (\ref{16}) we get $\delta T=0$.
However, if we keep the contribution of the fast oscillations, we
get a surface change of temperature 
\begin{equation}
\delta T\simeq\frac{2iT}{n\omega}\int dv\cdot vf_{s}(v)\cdot\delta(x)+...\:\:.\label{17}
\end{equation}
(\emph{i.e.}, $Re\left(\delta Te^{-i\omega t}\right)$). The $\delta$-type
contribution in equation (\ref{17}) corresponds to the surface sheath
in plasma heating models.\cite{key-6,key-25}

Similar calculations of the penetration depth can be made for a plasma
confined between two plane-parallel walls (or other geometries); the
result depends on the boundary conditions incorporated in parameters
like $f_{s}$.\cite{key-24} The boundary parameter $f_{s}$ is a
model parameter; we may take $f_{0}+f_{s}=0$ ($f(x=0,v)=0$) as a
natural assumption, an equation which provides the parameter $f_{s}$.
For $f_{s}=-f_{0}$ the field $E_{1}$ at the surface (maximum value)
is of the order $E_{1}\simeq E/\mid\varepsilon\mid$, where $E$ is
the internal uniform field given by equation (\ref{7}). The surface
change in temperature (equation (\ref{17})) can be written in this
case as 
\begin{equation}
\delta T=\frac{1}{2\pi}\left(\frac{E_{0}}{q/a^{2}}\right)T\cdot a\delta(x)\label{18}
\end{equation}
 (for $\omega\ll\omega_{0}$), where $a$ is the mean separation distance
between the particles ($a=n^{-1/3}$); $q/a^{2}\gg E_{0}$ is an electric
field of the order of the microscopic (inter-particle) field.

In conclusion, it is shown in this paper that the penetration of an
oscillating electric field in a semi-infinite classical plasma obeys
the standard exponential penetration law $e^{-x/\lambda_{e}}$ (beside
a uniform component), which may exhibit spatial oscillations, the
extinction length $\lambda_{e}$ (penetration depth, attenuation length)
being of the order $\lambda_{e}\simeq[\mid\varepsilon\mid/(1-\varepsilon)]^{1/3}v_{th}/\omega$;
($\varepsilon$ is the dielectric function, $\omega$ is the frequency
of the field and $v_{th}=\sqrt{T/m}$ is the thermal velocity). The
surface term is included explicitly in these calculations.

\textbf{Acknowledgments.} The author is indebted to the members of
the Laboratory of Theoretical Physics at Magurele-Bucharest for many
fruitful discussions. This work has been supported by the Scientific
Research Agency of the Romanian Government through Grants 04-ELI/2016
(Program 5/5.1/ELI-RO), PN 16 42 01 01/2016 and PN (ELI) 16 42 01
05/2016.

\end{document}